\documentclass[aps,prl,showpacs,preprintnumbers,twocolumn,superscriptaddress]{revtex4-1}
\usepackage{amsmath,amssymb}
\usepackage{bm}
\usepackage{tipa}
\usepackage{upgreek}
\usepackage{comment}
\usepackage{mathrsfs}
\usepackage{graphicx}
\usepackage{braket}
\usepackage{enumitem}
\usepackage{mathbbol}
\usepackage{gensymb}
\usepackage[normalem]{ulem}
\usepackage{color}
\usepackage[colorlinks,bookmarks=true,citecolor=blue,linkcolor=red,urlcolor=blue]{hyperref}
\usepackage{hyperref}
\usepackage{dsfont}

\begin{document}

\title{Entanglement smectic and stripe order}
\begin{abstract}
    Spontaneous symmetry breaking and more recently entanglement are  two cornerstones of quantum matter. We introduce the notion of anisotropic entanglement ordered phases, where the spatial profile of spin-pseudospin entanglement spontaneously lowers the four-fold rotational symmetry of the underlying crystal to a two-fold one, while the charge density retains the full symmetry. The resulting phases, which 
    we term \textit{entanglement smectic} and \textit{entanglement stripe}, exhibit a rich Goldstone mode spectrum and a set of phase transitions as a function of underlying anisotropies. We discuss experimental consequences of  such anisotropic entanglement phases  distinguishing them from more conventional charge or spin stripes.
    Our discussion of this interplay between entanglement and spontaneous symmetry breaking focuses on multicomponent quantum Hall systems realizing textured Wigner crystals, as may occur in graphene or possibly also in moir\'e systems, highlighting 
    the rich landscape and properties of possible entanglement ordered phases.
    
\end{abstract}

\author{Nilotpal Chakraborty}
\email{nilotpal@pks.mpg.de}
\affiliation{Max-Planck-Institut f\"{u}r Physik komplexer Systeme, N\"{o}thnitzer Stra\ss e 38, Dresden 01187, Germany}

\author{Roderich Moessner}
\affiliation{Max-Planck-Institut f\"{u}r Physik komplexer Systeme, N\"{o}thnitzer Stra\ss e 38, Dresden 01187, Germany}

\author{Benoit Doucot}
\affiliation{LPTHE, UMR 7589, CNRS and Sorbonne Universit\'e, 75252 Paris Cedex 05, France}

\maketitle

Spontaneous symmetry breaking forms much of the backbone of modern physics, particularly condensed matter \cite{anderson2018basic}. Besides the standard ideas of symmetry breaking in magnets and crystals, a remarkable development in the field over the last three decades has been the study of electronic phases in which the electronic density spontaneously breaks the underlying point group symmetry of the crystal. A variety of such symmetry broken phases have been intensively studied in quantum Hall systems and high-temperature superconductors, where they have been called, in analogy with the liquid crystal literature, electronic stripes, smectics and nematics \cite{Koulakovstripe,Moessnerstripe,emery1999stripe,Emerysmectic,Fradkinsmectic,lilly1999,du1999strongly,wu2011magnetic,Millistripe,fradkinreview,kivelsonreview}.

A more recent development in condensed matter has been the influence of ideas from quantum information, particularly the notion of entanglement \cite{nielsen2010quantum,horodecki2009quantum}. Entanglement between different degrees of freedom in a solid-state system offers the potential to realize interesting phases of matter and much has been done in reinterpreting known topological phases of matter through the entanglement lens. In this work we are particularly interested in entanglement between the different components of a multi-component electronic system, such as spin and pseudospin. More specifically, we focus on multi-component quantum Hall systems, which have witnessed a flurry of recent work both due to various spin-pseudospin entangled phases as well as connections with the physics of moir\'e systems \cite{Lianent,Murthyspin,stefanidis2023spin,Tarmagic,BultinckPRX}. 

In quantum Hall systems the energy bands form discrete Landau levels. We focus on the lowest Landau level which has four sub-levels labelled by the different combinations of spin and pseudospin. Such a situation occurs for the case of graphene where the pseudospin is the valley degree of freedom. On filling one of the four sub-levels, the ground state realizes a quantum Hall ferromagnet purely due to Coulomb interaction and at small doping away from unit filling the ground state is a multi-component skyrmion crystal, the charge density of  which has a discrete rotational symmetry \cite{Breyskyr,sondhiskyr}.
Due to the two components, the Coulomb interaction is almost $SU(4)$ symmetric when the lattice spacing is much smaller than the magnetic length \cite{goerbig2006electron,aliceaprb}. 
In this work, we show that weak $SU(4)$ symmetry breaking short-range interactions cause the spin-pseudospin entanglement profile to spontaneously break the discrete rotational symmetry of the charge density, giving rise to phases with anisotropic unidirectional entanglement - the \textit{entanglement smectic} and the \textit{entanglement stripe}.

\begin{figure*}
    \centering
    \includegraphics[scale  = 0.47]{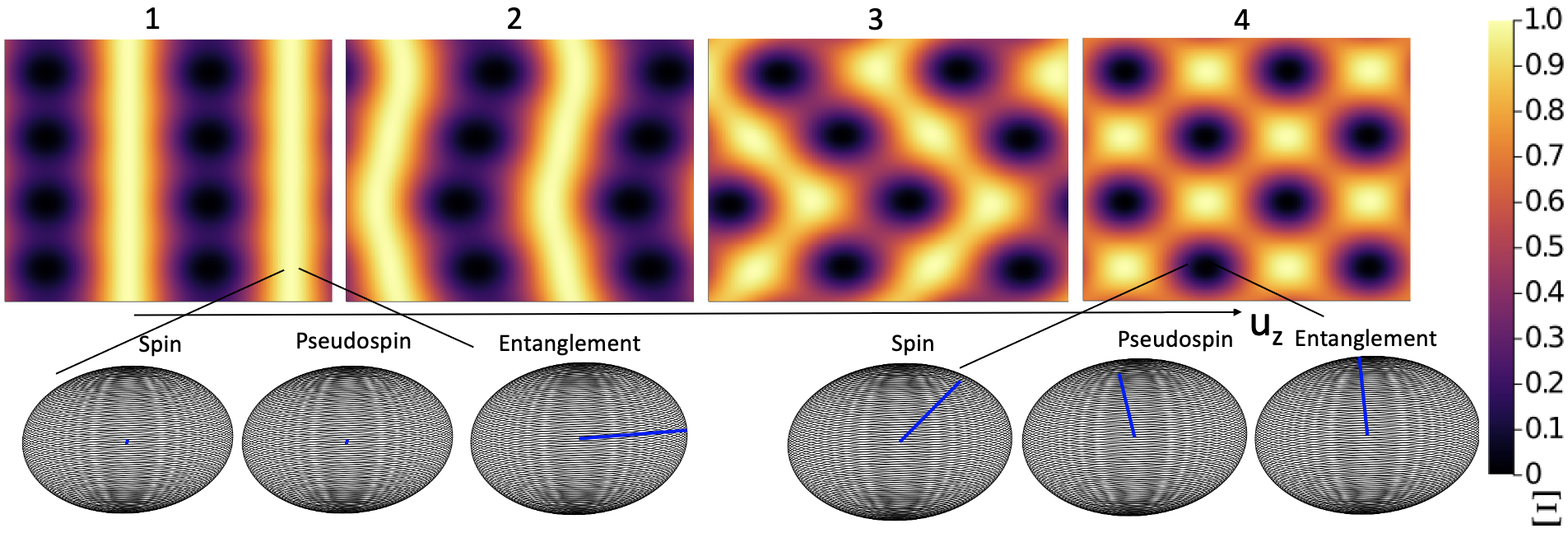}
    \caption{Entanglement patterns of the various entanglement ordered phases obtained on tuning the easy-axis anisotropy coupling constant ($u_z$ in Eq.(\ref{Eq. 2})) in the regime $J \gg g \gg \Delta$ (see eq. \ref{efunc} and the last paragraph in the model section). Top row: From left to right we see first, the formation of continuous straight lines of maximal entanglement $\Xi = 1$, along the $y-$axis which have a discrete periodicity along $x$ - the \textit{entanglement smectic} (1). Second, these straight lines start to modulate and form wavy patterns with modulated entanglement, which break the continuous translation symmetry of the smectic, while still maintaining a unidirectional nature - the \textit{entanglement stripe} (3). As $u_z$ gets closer to $u_p$, $C_4$ symmetry is restored (4). Importantly, all above patterns have the same $C_4$ symmetric charge density shown in Fig. \ref{Fig2} - the entanglement profile breaks $C_4$ down to $C_2$. See Fig. \ref{Fig3}a, to see $(u_z,u_p)$ values of 1-4. Bottom row: Bloch sphere representation of the local spin, pseudospin and entanglement vector for maximally (left) and minimally (right) entangled regions. }
    \label{Fig1}
\end{figure*}

Besides highlighting the \textit{entanglement smectic} and \textit{entanglement stripe} phases, we also show that there exists a cascade of phase transitions between such entanglement orders. These unidirectional phases, as a result of their entanglement structure have a complex spin (pseudospin) profile which could be detected in future STM experiments. Due to the broken approximate SU(4) symmetry we obtain
a rich low-lying mode spectrum  comprising  15 modes, out of which two remain gapless, and a phonon. Further, due to the broken discrete rotational symmetry, there is a large anisotropy in the dispersion of these low-lying modes resulting in a salient response of these entanglement ordered phases to magnon transport. Our work opens the possibility of realizing anisotropic entanglement ordered phases and entanglement liquid crystals.

\textit{\bf{Model:}}
For any quantum Hall ferromagnet, one can write the total energy for a 4-component spinor $\psi$ in an effective continuous theory as \cite{sondhiskyr, MoonPRB}
\begin{multline}
    E(\mathbf{\psi}) = J \int d^2\textbf{r} \bigg[ \dfrac{(\nabla \psi,\nabla \psi)}{(\psi,\psi)}-\dfrac{(\nabla \psi, \psi)(\psi,\nabla \psi)}{(\psi,\psi)} \bigg] + \\
    g\int V(\mathbf{r}-\mathbf{r'}) Q(\textbf{r})Q(\textbf{r'}) 
    d^2\textbf{r}\:d^2\textbf{r'} + \int E_A (\mathbf{r}) d^2 \mathbf{r}
    \label{efunc}
\end{multline}
where $J$ and $g$ are coupling constants for the exchange and topological charge terms (see eqs. 4 and 5 in supp mat for expressions for $Q$), $(a,b) \equiv \sum_{n=1}^4 a^*_n b_n$ is the complex inner product, $V(\bf{r} - \bf{r}')$ is the two body Coulomb potential (which we replace with a delta function term for analytical convenience in this work \cite{notsupp}) and $E_A$ is the anisotropy energy functional which breaks the approximate $SU(4)$ symmetry of the Coulomb interaction.
We use the energy functional given in \cite{kharithall,LianPRL} for anisotropies relevant to monolayer graphene, where pseudospin is the valley degree of freedom.
\begin{equation}
    E_A(\ket{\psi (\mathbf{r})}) = A \frac{\Delta_Z}{2}(u_p(M_{Px}^2 + M_{Py}^2) + u_zM_{Pz}^2 - M_{Sz})
    \label{Eq. 2}
\end{equation}
where $\Delta_Z \geq 0$ is proportional to the applied magnetic field, $u_p$ and $u_z$ characterize the pseudospin anisotropies which can be generated by short range interactions or lattice deformations due to phonon couplings and $A$ is a dimensionless constant. We include all terms which break the approximate SU(4) symmetry in $E_A$. Note that $E_A$ has an xy symmetry in plane.
\begin{figure*}
    \centering
    \includegraphics[scale  = 0.35]{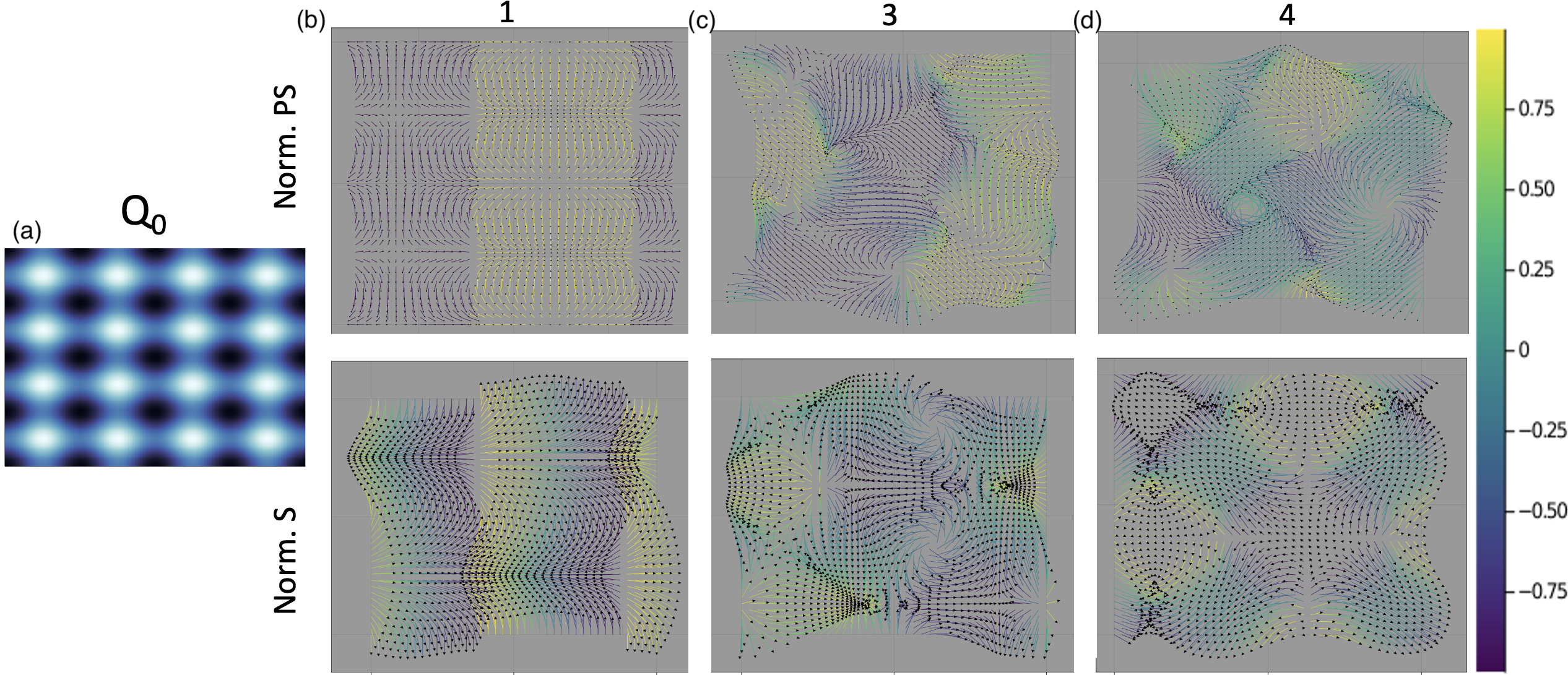}
    \caption{Rotationally invariant $(C_4)$ topological charge density (proportional to the charge density for quantum Hall skyrmion crystals) structure for the multi-component ($\mathbb{C}P(3)$) textured crystals and the normalized spin and pseudospin ($\bm{M}_{P(S)} / \rm{cos} \alpha$) profiles obtained by tuning $u_z$. The length and direction of the arrows indicate the in-plane component and the color represents the out-of-plane component - violet for the south pole and bright yellow for the north (same direction as $B$-field). Note that apparent discontinuities in the normalized spins are not real discontinuities since these regions are maximally entangled regions where the actual spin/pseudospin components vanish. The numbers on top of the plots correspond to $u_z$ values in Fig. \ref{Fig3} a.} 
    \label{Fig2}
\end{figure*}

To develop an analytical parameterization for skyrmion crystals for the above general energy functional, one must first consider an ansatz for  the  minimal energy configurations which imposes the constraint that the wavefunctions must be holomorphic. A natural choice of almost periodic holomorphic functions is the set of theta functions. The theta functions in our ansatz are 
\begin{equation}
    \theta_p(z) = \sum_{n = -\infty}^{\infty} e^{-\pi bd\frac{d}{a}(n+p/d)^2 + 2 \pi \frac{d}{a}(n + p/d)z} 
    \label{thetafunc}
\end{equation}
where $z = x + iy$ and we have $d$ skyrmions in a $a \times b$ unit cell ($d = 4$ and $a=b$ for this work) and $p \in (0,1,..,d-1)$ in accordance with the Riemann-Roch theorem \cite{debarre2005complex}. Our ansatz is for the simple case of a square lattice with periodic boundary conditions, however, symmetry breaking features of the entanglement measure presented in this work carry over to other shapes. 

Using this basis of theta functions we construct the holomorphic spinor $\ket{\psi(\mathbf{r})} = F \ket{\psi_0 (\mathbf{r})}$, where $\ket{\psi_0 (\mathbf{r})} = (\theta_0,\theta_1,\theta_2,\theta_3)^T$ and $F$ is a $4 \times 4$ $SU(4)$ matrix. The topological charge density term (see Methods) endows a stiffness to the crystal, and it was shown that all possible unitary matrices $F$ give rise to degenerate minimal energy configurations for skyrmion crystals with $SU(4)$-symmetric interactions~\cite{Dimaskyrmion}.  

However, the anisotropy term lifts this large degeneracy and selects a subset among the $SU(4)$ manifold of $F$ matrices. Our assumptions to allow for such an analytic skyrmion crystal ansatz imply that we are working in a regime with $J \gg g \gg \Delta_z$. To obtain a general $SU(4)$ matrix $F$, we develop a parametrization technique, inspired from random matrix theory \cite{diaconis2017hurwitz}, in terms of Euler angles (see Methods). Obtaining $\ket{\psi (\bf{r})}$ then reduces to a minimization problem for every $(u_p,u_z)$. 

\textit{\bf{Schmidt decomposition and entanglement:}}
Any normalized four-component complex ($\mathbb{C}P^3$) spinor can be written via a Schmidt decomposition in the form 
\begin{equation}
    \ket{\psi (\bf{r})} = \rm{cos}\frac{\alpha}{2}\ket{\Phi_P}\ket{\Phi_S} + e^{i\beta}\rm{sin}\frac{\alpha}{2} \ket{\Psi_P}\ket{\Psi_S}
    \label{schmidtdec}
\end{equation}
where $\ket{\Phi_{S(P)}}$,$\ket{\Psi_{S(P)}}$ are orthogonal basis vectors belonging to the spin (pseudospin) subspace (see \cite{doucotent} for details). These basis vectors are parameterized by the usual spherical angles $(\theta_{S(P)},\phi_{S(P)})$. Along with $(\alpha,\beta)$, these angles define three Bloch vectors for spin, pseudospin and entanglement. The Bloch vectors defining the local spin and pseudospin magnetizations can be obtained from the spinor directly via 
\begin{equation}
\begin{aligned}
    \mathbf{M}_S (\mathbf{r}) &= \braket{\psi(\mathbf{r})| 1 \otimes \bm{\sigma} | \psi(\mathbf{r})} = \hat{n}(\theta_S,\phi_S)\rm{cos}(\alpha) \\
    \mathbf{M}_P (\mathbf{r}) &= \braket{\psi(\mathbf{r})| \bm{\sigma} \otimes 1 | \psi(\mathbf{r})} = \hat{n}(\theta_P,\phi_P)\rm{cos}(\alpha)
    \end{aligned}
\end{equation}
where $\hat{n}$ is the unit vector on the sphere and $\bm{\sigma}$ is the vector comprising the Pauli matrices. In addition to these two, the entanglement Bloch vector is written as $\bm{m}_E = (\rm{sin}\alpha \, \rm{cos}\beta,\rm{sin}\alpha \, \rm{sin}\beta,\rm{cos}\alpha)$. 

An immediate consequence of the entanglement between spin and pseudospin is the respective Bloch vectors shortening and hence exploring the interior of the sphere. Particularly drastic situations arise when at certain points in space there is maximal entanglement, i.e, the $\bm{M}_{P(S)} = \bm{0}$   (for $\alpha = \pi/2$). A convenient entanglement measure, in this context, is $\Xi = \rm{sin}^2 \alpha$, since $\Xi = 0$ when there is no entanglement and $\Xi = 1$ for maximal entanglement \cite{doucotent}. We note that the entanglement measures for such multi-component spinor order parameters can differ slightly from that of the quantum state in a many-body Bose-Einstein condensate. We refer the reader to \cite{vedralodlro} for more details, our work focuses purely on spin-valley entanglement as defined in this section.

\textit{\bf{Entanglement smectics and stripes:}}
On analyzing the patterns of $\Xi (\bm{r})$ across the bulk, for certain cuts along $u_p-u_z$ parameter space we find phases with anisotropic profiles of spin-pseudospin entanglement. In these phases, the spatial profile of maximally entangled regions spontaneously there is no preference for $x$ or $y$-axis in any term in eq. \ref{efunc}) breaks the four-fold ($C_4$) rotational symmetry of the underlying charge density of the crystal (Fig. \ref{Fig2}), down to a two-fold one ($C_2$). 

We identify two such entanglement orders - the \textit{entanglement smectic} and \textit{entanglement stripe}. In the former, maximally entangled regions ($\Xi = 1$) form equidistant parallel lines as in Fig. \ref{Fig1}-1, whereas in latter, the continuous translational symmetry (along $y$) of the maximally entangled regions is absent but the entanglement profile still maintains a unidirectional, discrete rotational symmetry broken pattern as in Fig. \ref{Fig1}-3, much like uni-directional density waves.
The symmetry features of these entanglement ordered phases are similar to the electronic smectic/stripe phase, discussed in quantum Hall and high-$T_c$ literature \cite{Koulakovstripe,Moessnerstripe,Fradkinsmectic,emery1999stripe,Emerysmectic}. However, unlike charge or spin stripes (smectics) phases, the entanglement ordered phases have a four-fold symmetric charge density. Moreover, unlike spin stripes, the spins and pseudospins have textures along both directions in tne entanglement smectic and stripe, it is only the entanglement measure which has a one-dimensional nature.

Such entanglement orders arise due to competition between different anisotropy terms given in eq. (\ref{Eq. 2}) in the presence of a dominant Coulomb interaction. Conventionally, the presence of these anisotropies would orient the spin and pseudospin vectors either in-plane or out-of-plane. However, in the $J \gg g \gg \Delta_z$ regime, the dominant Coulomb term imposes a topological charge density profile, such as the square lattice profile in Fig. \ref{Fig2} captured by our ansatz in eq. (\ref{thetafunc}). As a consequence, we obtain a constrained optimization problem where the spins (pseudospins) cannot be all in-plane or out-of-plane and no single term in eq. (\ref{Eq. 2}) can be maximally satisfied. A compromise is allowed by spin-pseudospin entanglement which shortens the spin(pseudospin) vectors, and thereby reduces the energy loss from pseudospin terms in eq. (\ref{Eq. 2}). Such a reduction comes at a cost via a smaller gain from the Zeeman term which favours a larger spin magnitude. 

Such an optimization problem induces the following features: First, there is no reduced $SU(2)$ description of these orders in terms of a collection of spin/pseudospin or even entanglement skyrmions \cite{doucotent,LianPRL}, since the topological charge density constraint induces textures in all channels. Spin-pseudospin entanglement is maximal in regions which cost the most energy for a certain $(u_p,u_z)$. Second, the presence of a crystal adds the additional richness of packing of such non-uniform entanglement regions, especially for regions of maximal entanglement. These regions correspond to points at which the entanglement Bloch vector lies on the equator, which in turn corresponds to lines in real space (see Fig. \ref{Fig1}). Hence, the problem becomes one of the optimal arrangement of such lines, inspiring connections with liquid crystals.
\begin{figure}
    \centering
    \includegraphics[scale  = 0.4]{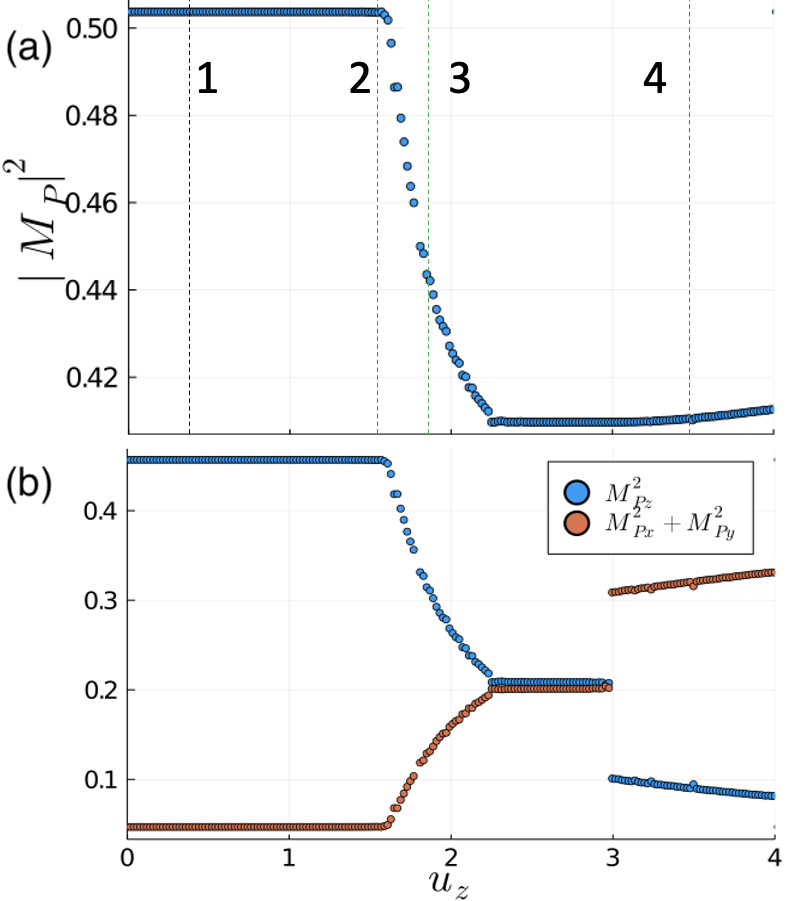}
    \caption{Entanglement order transitions ($u_p = 3\Delta_z$). (Top) Total pseudospin magnitude, the different numbered dash lines correspond to the respective profiles in Figs. \ref{Fig1} and \ref{Fig2}. A continuous transition at $u_z = u_p/2$, from the \textit{entanglement smectic} to the \textit{entanglement stripe}. See text for comments on other transitions. (Bottom) Out-of-plane and in-plane component of the pseudospin.}
    \label{Fig3}
\end{figure}

When $0 < u_z < u_p/2$, the maximal energy cost comes from the in-plane pseudospin component. Due to the fixed topological charge constraint all pseudospins cannot be put out-of-plane. Hence, entanglement is energetically favoured in regions with maximal in-plane pseudospin, as it reduces the associated anisotropy cost. As a consequence, we see in Fig. \ref{Fig3}(a), the entanglement stays constant till $u_z \sim u_p/2$, and the net out-of-plane pseudospin is much higher than that in-plane.    Further, as we see from the spin profile, since $u_p > \Delta_z$ maximally entangled columns include spins pointing in opposite directions since the system compromises its gain from the zeeman term to prioritize reducing the in-plane pseudospin anisotropic cost (see Fig. \ref{Fig1}-1 and Fig. \ref{Fig2}(a)).
To maximize the area covered by the lines of maximal entanglement (by avoiding intersections) while respecting the symmetries imposed by the dominant Coulomb interaction which imposes square lattice charge density, the lines must be parallel. The system spontaneously chooses one of the two axes to align these lines with, thereby breaking the four-fold ($C_4$) symmetry down to two-fold ($C_2$). 

On increasing $u_z$ beyond $u_p/2$, the anisotropy costs increase and so does the entanglement (see Fig. \ref{Fig3}(a)) since the easy-axis anisotropic cost becomes comparable to the easy-plane cost. Moreover, the distribution of maximally entangled regions changes as $u_z$ approaches $u_p$, resulting in transitions between different entanglement orders. Due to the competition between minimizing easy-axis and easy-plane costs, the entanglement spreads out spatially and the uniformity along one-dimensional smectics breaks up into a more general stripe order (see \ref{Fig1}-3). Eventually when $u_z \sim u_p > \Delta_z$, the system restores four-fold rotational ($C_4$) symmetry (see \ref{Fig1}-4). In the $u_z > u_p > \Delta_z$ regime, the pseudospin is mostly in-plane (see Fig. \ref{Fig2}(d) and Fig. \ref{Fig3}(b)), and in regions where the topological charge density forces it to go out of plane, such as the core of the topological defects, the system maximally entangles, as seen in Fig. \ref{Fig2}(d).  

\textit{\bf{Entanglement order transitions:}}
As a consequence of such delicate competition we have two continuous entanglement order transitions. To study the nature of these transitions, we plot the average pseudospin magnitude and the out-of-plane component summed over all sites in Fig. \ref{Fig3}. First there is a continuous transition at $u_z = u_p/2$ which corresponds to the deviation of maximally entangled regions from straight lines into wavy lines and modulated entanglement along those lines, thereby breaking the continuous translational symmetry. An appropriate order parameter for such a transition would be the amplitude of the periodic undulation of these lines along the transverse ($x$) direction. This corresponds to the \textit{entanglement smectic} to \textit{entanglement stripe} transition. Next, there is another continuous transition corresponding to the restoration of $C_4$ symmetry. Such a transition occurs via melting of the \textit{entanglement stripe}. Finally as we see from the jump in the in and out-of-plane components in Fig. \ref{Fig3}b, there is another transition at $u_p = u_z$, however, the symmetries of the entanglement structure remain the same. 

\textit{\bf{Entanglement as angle-dependant impedance to magnon transport:}}
When $J$ is the only non-zero energy scale, we get a series of effective Landau  levels with the lowest one pinned to zero energy - the \textit{Riemann-Goldstone Landau level} \cite{Chakrabortymagnon}. For $g 
\neq 0$, the three Goldstone modes for the isotropic $SU(2)$ case acquire dispersion. In this work, for $SU(4)$, we demonstrate the existence of 15 Goldstone modes and a single phonon mode in the Riemann-Goldstone Landau level based on counting independent small deformations in the zero momentum sector \cite{notsupp}. Using the axial gauge and a real-space discretization scheme \cite{notsupp}, we  obtain their dispersion. 

The anisotropy breaks the approximate $SU(4)$ symmetry and gaps out 13 of the 15 modes. However, in addition to the phonon (only gapless in the continuum), there are remnant $U(1)_P$ and $U(1)_S$ symmetries which result in two gapless modes corresponding to a rotation about the $z-$axis of both spin and pseudospin.

A salient feature due to the highly anisotropic nature of the entanglement profile in the various phases, is the considerable anisotropy in the dispersion relation. One particularly notable case is that for the \textit{entanglement smectic}. The group velocities for the two Goldstone modes of the \textit{entanglement smectic} are heavily suppressed along one axis ($q_x$) of the Brillouin zone as compared to the other ($q_y$) (see Fig. S1). This suppression results in striking experimental consequences for magnon transport in quantum Hall heterojunction experiments.

Magnon transport in quantum Hall heterojunctions for graphene has emerged as a promising platform to detect the underlying spin structure of the quantum Hall phase diagram \cite{wei2018electrical,zhou2020solids,zhou2022strong,zhou2021magnon,assouline2021excitonic}. An intriguing question arising from our work is if mesoscopic magnon transport can diagnose entanglement. Say for example, we have the \textit{entanglement smectic} with uniform columns across which the spin vanishes (due to maximal entanglement). Such a configuration will not allow any low energy spin-wave approaching at normal incidence to pass through as the columns will act as a blockade, as seen from the extremely flat dispersions of the gapless modes along $q_x$ in Fig. S1. Hence, there will be almost full reflection of any such incoming magnon except perhaps at very special energies. Moreover, due to the strong anisotropy, there will be a strong dependence of the transmission on the incidence angle of the magnon. Such a drastic impedance mismatch is a unique consequence of such entanglement smectics, absent from any other kind of electronic order. 

\textit{\bf{Discussion:}}
We have introduced anisotropic entanglement ordered phases which spontaneously break the discrete rotational symmetry of the charge density of the underlying crystal- the \textit{entanglement smectic} and the \textit{entanglement stripe}.These phases have salient magnon transport features observable in ongoing experiments.
These phases also have distinct spatial profiles of the effective spin (pseudospin) magnetization which could be detected via scanning tunneling microscopy given the recent success in detecting symmetry broken phases in the lowest Landau level of graphene \cite{feldman2016observation,liu2022visualizing}.

Our work, given the analogy with earlier seminal works in the quantum Hall and high-$T_c$ literature \cite{Koulakovstripe,Moessnerstripe,fradkinreview,Fradkinsmectic,Emerysmectic,emery1999stripe,kivelsonreview,Millistripe}, introduces the theoretical framework of \emph{entanglement liquid crystals} and raises the intriguing possibility of an \textit{entanglement nematic} phase of matter, akin to the nematics found in liquid crystals. Furthermore, several connections have been proposed between the physics of twisted bilayer graphene \cite{cao2018correlated,cao2018unconventional} and moir\'e systems in general, to multicomponent quantum Hall physics and skyrmions \cite{BultinckPRX,khalaf2021charged,khalaf2022baby}. Such connections raise the exciting possibility of studying variations of such entanglement ordered phases between the different flavours in these systems, especially the dichalcogenides \cite{seyler2019signatures,wang2020correlated}. Moreover, viewing quantum matter with such enlarged phase-spaces through the lens of entanglement also raises the intriguing possibilites of physics related to the entanglement phase $\beta$. There exists a poisson bracket structure between $\alpha$ and $\beta$ in the six-dimensional phase-space associated with $\mathbb{C}P(3)$, hence this conjugate phase could result in possible entanglement luttinger liquids and also interesting qubit architectures.

 \textit{\bf{Acknowledgements:}}
This work was in part supported by the Deutsche Forschungsgemeinschaft under grants SFB 1143 (project-id 247310070) and the cluster of excellence ct.qmat (EXC 2147, project-id 390858490). B. D. thanks MPIPKS for its generous hospitality during several extended visits, which were crucial for the realization of this project. This research was supported in part by grants NSF PHY-1748958 and PHY-2309135 to the Kavli Institute for Theoretical Physics (KITP).

\bibliography{Bib.bib}


\title{Supplementary material for "Entanglement stripe and smectic order"}




\onecolumngrid

\tableofcontents

\section{Discretizing $\mathbb{C}P(3)$ energy functional.}

In this work we consider a delta function Coulomb potential and a reference topological charge density configuration generated by theta-functions, which is set to be the minimum of the energy functional (see discussion in \cite{Chakrabortymagnon} for experimental relevance).

A point in $\mathbb{C}P(N-1)$ is a line in $\mathbb{C}^N$ spanned by a non-zero $N$-component vector denoted by $\ket{\psi}$. As a preliminary step to discretize the energy functional let us specify the geodesics in $\mathbb{C}P(N-1)$ between $\mathbb{C}\ket{\psi_1}$ and $\mathbb{C}\ket{\psi_2}$ ($\mathbb{C} \ket{\psi}$ is the complex line in $\mathbb{C}^N$ space going through the vector $\psi$ and through the origin) 
with the $\braket{\psi|\psi} =  1$ constraint. We consider a map $\lambda \rightarrow \ket{\psi(\lambda)}$ such that $\ket{\psi (\lambda = 0)} = \ket{\psi_1}$, $\ket{\psi (\lambda = 1)} = \ket{\psi_2}$ and $\braket{\psi(\lambda)|\psi(\lambda)} =  1$. We may further choose phases of $\ket{\psi_1}$ and $\ket{\psi_2}$ such that $\braket{\psi_1|\psi_2} \in \mathbb{R}$. We can define the angle $\theta \in [0,\pi/2]$ such that $\braket{\psi_1|\psi_2} = \rm{cos}\, \theta$. The geodesic then corresponds to rotating $\ket{\psi_1}$ into $\ket{\psi_2}$ with constant angular velocity, which can be written as 
 
 \begin{equation}
     \ket{\psi(\lambda)} = \dfrac{\rm{sin}(\theta (1-\lambda)) \ket{\psi_1} + \rm{sin}(\theta \lambda)\ket{\psi_2}}{\rm{sin}\, \theta} 
 \end{equation}
 The geodesic length in  $\mathbb{C}P(N-1)$ is simply $\theta$, which in the general case of $\braket{\psi_1|\psi_2} \in \mathbb{C}$ is given by $\rm{cos}\, \theta = (\braket{\psi_1|\psi_2}\braket{\psi_2|\psi_1})/(\braket{\psi_1|\psi_1}\braket{\psi_2|\psi_2})^{1/2}$. 

 Say $\ket{\psi_1}$ and $\ket{\psi_2}$ correspond to the spinors on neighbouring sites, which implies that they differ by a small angle $\theta \approx 0$, we can write 
 \begin{equation}
     \theta^2 = 1 - \dfrac{\braket{\psi_1|\psi_2}\braket{\psi_2|\psi_1}}{\braket{\psi_1|\psi_1}\braket{\psi_2|\psi_2}}
  \end{equation}
 which is gauge invariant. Therefore a natural discretization for the local exchange term can be written as 
 \begin{equation}
     E_{\rm{exc}} = J \sum_{<i,j>}  \bigg(1- \dfrac{\braket{\psi_i|\psi_j}\braket{\psi_j|\psi_i}}{\braket{\psi_i|\psi_i}\braket{\psi_j|\psi_j}}\bigg) 
 \end{equation}
 where the sum is over nearest neighbour  pairs.

 To discretize the topological charge density terms, one needs to consider 4 spinors associated with the sites of a plaquette in real space. The procedure to calculate this for the simpler $SU(2)$ skyrmion crystal using local spin vectors was introduced in ref.\cite{Chakrabortymagnon}. The basic idea is to consider the nearest neighbours pairwise and calculate the topological charge between two nearby geodesics. We show here how to do this for the $\mathbb{C}P(3)$ manifold.

Consider a closed path $\gamma$ in $\mathbb{C}P(N-1)$ parametrized by $t \in [0,1]$. The topological charge associated to this path has the general form $\oint \mathcal{A}$, where $\mathcal{A}$ is a 1-form on $\mathbb{C}P(N-1)$ whose curl is equal to the canonical
symplectic two form on $\mathbb{C}P(N-1)$ associated to the Fubini-Study metric. A convenient way to evaluate $\oint \mathcal{A}$ is to take a map $t \rightarrow \ket{\psi(t)} \in \mathbb{C}^N-{0}$ such that $\mathbb{C} \ket{\psi(t)} = \gamma(t)$ at all $t$. Such a procedure is called \textit{lifting} the path. It is essential that this \textit{lifted} path be closed as well, i.e $\ket{\psi(0)} = \ket{\psi(1)}$, in order to get an expression which is independent of the choice of such a path. Then we get
\begin{equation}
    Q_{\gamma} = \frac{1}{2i} \int_{0}^{1} \dfrac{\braket{\psi|\dot{\psi}} - \braket{\dot{\psi}|\psi}}{\braket{\psi|\psi}}dt
\end{equation}

Let us now consider a geodesic quadrangle on $\mathbb{C}P(N-1)$ where $\mathbb{C}\ket{\psi_j'}$ is very close to $\mathbb{C}\ket{\psi_j}$ ($j = 0,1$), so we may set $\ket{\psi_j'} =\ket{\psi_j} + \ket{\chi_j}$. If we consider the quadrangle comprising the two initial ($\psi_0,\psi_1$) and two final ($\psi'_0,\psi'_1$) points, we can write
\begin{equation}
\begin{split}
      Q_{01} &= \rm{arg}\bigg(\dfrac{\braket{\psi_0|\psi'_0}}{\braket{\psi_0|\psi_0}}\bigg) + \rm{arg}\bigg(\dfrac{\braket{\psi_0'|\psi_1'}}{\braket{\psi_0|\psi_1}}\bigg) + \rm{arg}\bigg(\dfrac{\braket{\psi_1'|\psi_1}}{\braket{\psi_1|\psi_1}}\bigg)\\
    &= \frac{1}{2i}\bigg(\dfrac{\braket{\psi_0|\chi_0} - \braket{\chi_0|\psi_0}}{\braket{\psi_0|\psi_0}} + \dfrac{\braket{\psi_0|\chi_1} + \braket{\chi_0|\psi_1}}{\braket{\psi_0|\psi_1}} - 
     \quad \dfrac{\braket{\chi_1|\psi_0} + \braket{\psi_1|\chi_0}}{\braket{\psi_1|\psi_0}} + \dfrac{\braket{\chi_1|\psi_1} - \braket{\psi_1|\chi_1}}{\braket{\psi_1|\psi_1}}\bigg)
\end{split}
\end{equation}
The total contribution from the plaquette $0123$ is then given by 
$Q_{0123}=Q_{01}+Q_{12}+Q_{23}+Q_{30}$. Finally, the contribution to the topological charge density term for the site $0$ is given by $\sum_{0 \in \square} Q_{\square}^2$.

 \section{Euler angle parameterization of a general SU(4) matrix.}
 Let $(e_1,e_2,...,e_N)$ be an orthonormal basis in $\mathbb{C}^N$. The idea is to 
 use a recursive procedure in order to parametrize all matrices in $SU(N)$ corresponding to isometries $f:\mathbb{C}^N \rightarrow \mathbb{C}^N$. 
 Suppose that we know how to find a rotation $g_1 \in SU(N)$ such that $g_1(e_1) = f(e_1)$,
 which amounts to saying that the first columns of the matrices representing $f$ and
 $g_1$ are identical. The explicit parametrization of such a $g_1$ will be given below. Let us introduce a rotation $f_2$ given by $f = g_1f_2$, then $f_2(e_1) = e_1$. The matrix representing $f_2$ has $1$ in the top left and then a diagonal block $M_2 \in SU(N-1)$. We apply the same procedure to $M_2$, i.e we search for $g_2 \in SU(N-1)$ such that $g_2(e_1) = f_2(e_1) = e_1$ and $g_2(e_2) = f_2(e_2) \in \rm{Vect}(e_2,e_3,...,e_N)$. Then $f_2 = g_2f_3$ with $f_3(e_1) = e_1$ and $f_3 (e_2) = e_2$. Now $f_3$ is a matrix with two diagonal blocks, the identity matrix $\mathbb{1}_2$ and $M_3 \in SU(N-2)$. We keep applying this procedure to the nested sequence of groups $SU(N) \supset SU(N-1) \supset ... \supset SU(2)$. In this decreasing sequence of groups, $SU(N-p)$ is understood as the subgroup of $SU(N)$ of rotations which leave $e_1,...,e_p$ invariant. In the end we get
 \begin{equation}
     f = g_1 g_2 \dots g_{N-1}
     \label{fmat}
 \end{equation}

It remains then to find a convenient parametrization for the $g_1$ rotation. We know that $g(e_1) \in \rm{Vect}({e_1,..e_N})$ is given (where Vect$(e_1,\dots,e_i)$ implies that the vector has $i$ non-zero components). The idea is to go from $e_1$ to $g(e_1)$ step by step, switching on sequentially the non-zero component of $g(e_1)$ starting from $x_2$, then $x_3$ and so on up to $x_N$. For this, we write $f(e_1) = h_N h_{N-1} \dots h_2(e_1)$ with 
 \begin{equation}
 \begin{split}
     h_2(e_1) &\equiv v_2 \in \rm{Vect}(e_1,e_2) \\
     h_3(v_2) &\equiv v_3 \in \rm{Vect}(e_1,e_2,e_3)\\
     &. \\
     &. \\
     h_N(v_{N-1}) &\equiv g(e_1) \in \rm{Vect}(e_1,e_2,...,e_N)
     \label{seqrot}
 \end{split}
 \end{equation}
 Let us define $R_j(\theta_j,\phi_j,\psi_j) \in SU(N)$ as the rotation such that 
 \begin{equation}
 \begin{split}
     R_j(\theta_j,\phi_j,\psi_j)(e_k) &= e_k ; k \notin {j-1,j}\\
     R_j(\theta_j,\phi_j,\psi_j)(e_{j-1}) &= \rm{cos} (\theta_j) e^{i\psi_j} e_{j-1} + \rm{sin}(\theta_j)e^{i\phi_j}e_j \\
     R_j(\theta_j,\phi_j,\psi_j)(e_j) &= -\rm{sin} (\theta_j) e^{-i\phi_j} e_{j-1} + \rm{cos}(\theta_j)e^{-i\psi_j}e_j \\
 \end{split}
 \end{equation}
 
 Using such a parametrization one can show that  Eq.~(\ref{seqrot}) is always satisfied  with rotation matrices chosen as $h_j = R_j(\theta_j,\phi_j,\psi_j)$. We start from $h_N$, and we write $f(e_1) = g(e_1) = x_1e_1 + \dots + x_Ne_N$ ($\sum_i |x_i|^2 = 1$). We wish to find $v_{N-1} = y_1e_1 + \dots + y_{N-1}e_{N-1}$ ($\sum_i |y_i|^2 = 1$ and we can always choose $y_{N-1}$ to be a non-negative real number) and $\theta_N \in [0,\pi/2]$, $\phi_N \in [0,2\pi]$, $\psi_N \in [0,2\pi]$ such that $R_N (\theta_N,\phi_N,\psi_N)(v_{N-1}) = g(e_1)$. This condition implies that $y_k = x_k$ for $k = (1,...,N-2)$ and 
 \begin{equation}
 \begin{split}
     y_{N-1}\cos(\theta_N)e^{i\psi_N} &= x_{N-1}\\
     y_{N-1}\sin(\theta_N)e^{i\phi_N} &= x_N
     \label{receq}
\end{split}
 \end{equation}
 One can write the polar decomposition of the R.H.S of the above equation to obtain the three Euler angles. Once $v_{N-1} \in \rm{Vect}(e_1,...,e_{N-1})$ is determined we set to find $v_{N-2} \in \rm{Vect}(e_1,..,e_{N-2})$ and $[\theta_{N-1},\phi_{N-1},\psi_{N-1}]$ such that $R_{N-1} (\theta_{N-1},\phi_{N-1},\psi_{N-1})(v_{N-2}) = v_{N-1}$. 
 One can solve this using the similar procedure which resulted in Eq. (\ref{receq}). 
 Because $y_{N-1}$ is real and non-negative, we see that if we take the component of $v_{N-2}$ along $e_{N-2}$ to be also real and non-negative, then $\phi_{N-1}=0$.
 Continuing this process recursively one determines all the Euler angles, from $N-1$ to $2$. Therefore we can write
 \begin{equation}
     g_1 = R_N(\theta_N)R_{N-1}(\theta_{N-1})\dots R_2(\theta_2)
 \end{equation}
 where we have hidden the dependence on the other $\phi$ and $\psi$ angles for notational convenience. As we saw earlier, only $\phi_{N} \neq 0$, so $g_1$ in the above equation depends
 only on $2N-1$ real parameters, as it should because $f(e_1)$ belongs to the unit sphere in $\mathbb{C}^N$, whose dimension
 is precisely $2N-1$. Since we have to also decompose $g_2,\dots,g_{N-1}$ in a similar fashion, we have to introduce $2N-1+\dots+5+3 = N^2-1$ Euler angles for an arbitrary matrix in $SU(N)$.
 
 One can write these product decompositions as
 \begin{equation}
 \begin{split}
     g_1 &= R_N(\theta_{1,N})R_{N-1}(\theta_{1,N-1})\dots R_2(\theta_{1,2})\\
     g_2 &=  R_N(\theta_{2,N})R_{N-1}(\theta_{2,N-1})\dots R_3(\theta_{2,3})\\
     & . \\
     & . \\
     g_j &=  R_N(\theta_{j,N})R_{N-1}(\theta_{j,N-1})\dots R_{j+1}(\theta_{j,j+1})\\
     &.\\
     &.\\
     g_{N-1} &= R_N(\theta_{N-1,N})
 \end{split}
 \end{equation}
Using these results for eq. (\ref{fmat}) we obtain the matrix $f$.

\section{$\mathbb{C}P(3)$ equations of motion.}
The kinetic term of the Lagrangian gives the Berry phase and is written as:
\begin{equation}
    \mathcal{L}_k = \frac{1}{2i}\dfrac{\braket{\psi|\dot{\psi}} - \braket{\dot{\psi}|{\psi}}}{\braket{\psi|\psi}}
\end{equation}
Under a gauge transformation $\ket{\psi(t)} = f(t)e^{i\alpha(t)}\ket{\chi(t)}$, which gives
\begin{equation}
\mathcal{L}_k (\psi,\dot{\psi}) = \mathcal{L}_k(\chi,\dot{\chi}) + \dot{\alpha}
\end{equation}
 So provided $\alpha(0) = \alpha(1)$, we have $\int_0^1 \mathcal{L}_k (\psi,\dot{\psi}) dt = \int_0^1 \mathcal{L}_k (\chi,\dot{\chi}) dt $ (this in particular the case if we keep the end points of the path fixed, i.e $\alpha(0) = \alpha(1) = 0$). The potential term $\mathcal{L}_p(\psi^*,\psi)$ is invariant under gauge transformations $\psi_a \rightarrow \epsilon \psi_a $, with ($1 \leq a \leq N$) and $\epsilon$ an arbitrary complex number. From this we get   
 \begin{equation}
     \sum_{a=1}^{N} \frac{\partial \mathcal{L}_p}{\partial \psi_a}(\psi^*,\psi)\psi_a = \sum_{a=1}^{N} \frac{\partial \mathcal{L}_p}{\partial \psi^*_a}(\psi^*,\psi)\psi^*_a = 0
     \label{potgauge}
 \end{equation}
 The Euler-Lagrange equations of motion read
 \begin{equation}
     i \big( \dfrac{\dot{\psi}_a}{\braket{\psi|\psi}} - \dfrac{\braket{\psi|\dot{\psi}}}{\braket{\psi|\psi}^2}\psi_a\big) = \dfrac{\partial \mathcal{L}_p}{\partial  \psi_a^*}
 \end{equation}
 Multiplying the L.H.S by $\psi_a^*$ and summing over $a$ gives 0, which is compatible with the gauge invariance of the potential term as expressed in Eq. (\ref{potgauge}). Let us assign $s_a \equiv -i (\partial \mathcal{L}_p/\partial \psi_a^*)$, the above equation can then be written in terms of $s_a$ and it determines $\ket{\dot{\psi}}$ up to a vector proportional to $\ket{\psi}$. We can write
 \begin{equation}
     \ket{\dot{\psi}} = \braket{\psi|\psi} \ket{s} + b\ket{\psi}
 \end{equation}
 where $b$ is an arbitrary complex number. We can eliminate this redundancy in the choice of $b$ u\rm{sin}g the axial gauge. We wish to impose
 \begin{equation}
     \braket{g|\dot{\psi}} = 0
 \end{equation}
 where $g$ is a given time-independent $N$-component complex vector. Most often $g$ will be chosen of the form $g_a = \delta_{ai}$ with fixed $i$ (at a given site). Imposing the above determines $b$ provided that $\braket{g|\psi} \neq 0$. Then 
 \begin{align}
     b &= - \braket{\psi|\psi} \dfrac{\braket{g|s}}{\braket{g|\psi}}\\
     \ket{\dot{\psi}} &= \braket{\psi|\psi}\left(\ket{s} - \dfrac{\braket{g|s}}{\braket{g|\psi}}\ket{\psi}\right)
\end{align}

To obtain the linearized equations of motion we start from of variational ground-state $\ket{\psi}$ and look for time dependent solutions of the form $\ket{\psi} + \ket{\chi(t)}$ with $\ket{\chi(t)}$ small. To first order in $\chi$ we get
\begin{align}
    \dfrac{\ket{\dot{\chi}}}{\braket{\psi|\psi}} &- \dfrac{\braket{\psi|\dot{\psi}}}{\braket{\psi|\psi}^2}\ket{\psi} = \ket{s} \\
    s_a &= -i \sum_{b = 1}^{N}  \big( \dfrac{\partial^2 \mathcal{L}_p}{\partial \chi_a^* \partial \chi_b}\chi_b + \dfrac{\partial^2 \mathcal{L}_p}{\partial \chi_a^* \partial \chi_b^*}\chi^*_b \big)
\end{align}
 The above equation however does not have a solution when $\ket{\psi}$ is not an exact energy extremum, as it is the case in our variational procedure, where the energy functional is minimized over the \emph{global} choice of an $SU(4)$ matrix $F$.
 To rectify this we can project $\ket{s}$ into the subspace orthogonal to $\ket{\psi}$. This gives us the following form for the equations of motion
 \begin{equation}
     \ket{\dot{\chi}} = \braket{\psi|\psi}\big(\ket{s} - \dfrac{\braket{g|s}}{\braket{g|\psi}}\ket{\psi}\big)
 \end{equation}
However, to ensure discrete translational invariance one has to make a small modification and replace $\ket{\chi}$ by $\ket{w} = \langle g | \psi \rangle^{-1} \ket{\chi}$. The final equations of motion are
 \begin{equation}
     \ket{\dot{w}} = \dfrac{\braket{\psi|\psi}}{\langle g | \psi_0 \rangle}\left(\ket{s} - \dfrac{\braket{g|s}}{\braket{g|\psi}}\ket{\psi}\right)
 \end{equation}
where $\ket{s}$ in the R.H.S is now expressed as a function of $\ket{w}$.

Using this discretization scheme, we numerically obtain the dispersion of the Goldstone models for the entanglement smectic as shown in Fig. S1. Such anistropy leads to drastic consquences for magnon transport (see main text).

\begin{figure}
    \centering
    \includegraphics[scale  = 0.3]{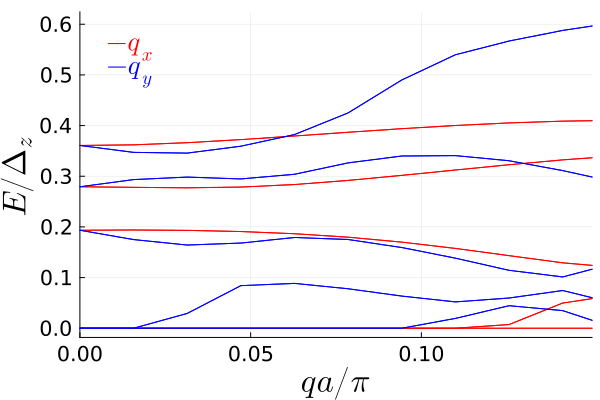}
    \caption{Lowest five modes' anisotropic dispersion of the entanglement smectic. Red (Blue) lines indicate dispersion along x (y) axis of first Brillouin zone. For weak values of $g$ ($g = J/20$, $u_p = 3 \Delta_z$ and $u_z = \Delta_z/10 $ here), the two gapless modes remain relatively flat along $q_x$ whereas they become dispersive along $q_y$. For higher values the two modes will eventually acquire a dispersion, however, the dispersion along $q_x$ will be strongly suppressed.
    }
    \label{Fig4}
\end{figure}

\section{Goldstone mode counting.}
In this section we present a mathematically rigorous argument for the existence of 15 type-preserving Goldstone modes (of which 13 are gapped out in presence of anisotropies) in the Riemann-Goldstone Landau level as well as a phonon, for the model that we considered above.

Let us examine the structure of small deformations at $\bm{q} = 0$. Let us consider the texture described by the map $z \rightarrow (\theta_0(z),\theta_1(z), \dots,\theta_{N-1}(z)) \equiv \ket{\psi(z)}$. We have two kinds of deformations in the $\bm{q} = 0$ sector around such optimal crystals.

a) \emph{Type-preserving} deformations: We deform $\ket{\psi(z)}$ into $\ket{\psi(z)}$ + $\ket{\chi(z)}$, with $\ket{\chi (z)}$ of the form 
\begin{equation}
    \chi_a(z) = \sum_{b= 0}^{N-1} M_{ab}\theta_b(z)
\end{equation}
where $M$ is a $N \times N$ matrix. We call such deformations \emph{type-preserving}, because they don't require to modify the theta function basis used to describe the deformed texture.
This gives us a continuous family of $q = 0$ deformations with $2(N^2-1)$ real parameters, after excluding gauge transformations corresponding to $M$ matrices proportional to the identity. 

b) Phonon deformations: These are of the form
\begin{equation}
    \chi_a(z) = u \psi'_a(z); \; u \in \mathbb{C}
\end{equation}

where $\psi'_a(z) = \partial \psi_a/\partial z$. Such deformations are not type-preserving since the derivative of a theta function doesn't
transform under lattice translations as a theta function, hence these continuous translations cannot be accounted for without changing the underlying theta function basis.
It is well known in the mathematics literature that line bundles (and their associated theta functions)
of fixed degree $N$ on a torus have exactly one free complex deformation parameter, which can be
visualized as the center of mass position of the set of zeros of theta functions
inside a given unit cell. So, this suggests that phonon deformations exhaust all type deforming deformations which preserve the discrete lattice periodicity of the texture.

A general $q = 0$ deformation may be written as
\begin{equation}
    \chi_a(z) = \sum_{b= 0}^{N-1} M_{ab}\theta_b(z) + u \psi'_a(z)
\end{equation}
with $\rm{trace}(M) = 0$. Such deformations involve $2N^2$ real parameters.

$SU(N)$ symmetry generates 15 independent flat directions, i.e, small deformations that do not change the  energy. We show below that these 15 directions are mutually orthogonal for the symplectic form that generates the linearized equations of motion. As a result we'll see that the 15 flat directions give 15 zero eigenvectors and 15 two-dimensional Jordan blocks. Recall that a 2D Jordan block arises from linearizing the free particle motion,
generated by $H=p^2/2$, in the vicinity of any equilibrium point $(p,q)=(0,q_0)$ in phase-space.

\textit{Symplectic form for small deformations in the $q = 0$ sector}: Take $\chi_a^{(1)} (z)$ and $\chi_a^{(2)} (z)$ to be two small deformations around the optimal Skyrmion crystal described by the holomorphic map $z \rightarrow \ket{\psi(z)}$. We define
\begin{equation}
    \Omega\big[\chi^{(1)}, \; \chi^{(2)}\big] \equiv i \int dxdy \bigg[ \dfrac{\langle \chi_2|\; (\mathbb{1} - P_\psi) \; |\chi_1 \rangle}{\langle \psi | \psi \rangle} -
    \dfrac{\langle \chi_1|\; (\mathbb{1} - P_\psi) \; |\chi_2 \rangle}{\langle \psi | \psi \rangle}\bigg]
\end{equation}
where $P_\psi = \ket{\psi(z)}\bra{\psi(z)}/\langle \psi(z) | \psi(z) \rangle$ and we  have hidden the $z-$dependence for notational convenience. We use this symplectic form because equations of the motion have the 
generalized Hamiltonian form:
\begin{equation}
    i\bigg( \dfrac{\dot{\psi}_a(\bm{r})}{\langle \psi (\bm{r}) | \psi(\bm{r}) \rangle} - \dfrac{\langle \psi(\bm{r}) | \dot{\psi}(\bm{r}) \rangle \psi_a(\bm{r})}{\langle \psi (\bm{r}) | \psi(\bm{r}) \rangle^2}\bigg) = \dfrac{\partial \mathbb{L}_p}{\partial \psi^*_a(\bm{r})}
\end{equation}
where $\mathbb{L}_p$ is the potential energy functional, and one can obtain these equations by imposing that
\begin{equation}
    \Omega \big[\dot{\psi},\chi ] = \int d^2 r \bigg(\dfrac{\partial \mathbb{L}_p}{\partial \psi^*_a(\bm{r})} \chi_a^*(\bm{r}) + \dfrac{\partial \mathbb{L}_p}{\partial \psi_a(\bm{r})}\chi_a(\bm{r})\bigg)
\end{equation}
for \textbf{all} possible small deformations $\chi_a(\bm{r})$, $1 \leq a \leq N$. $\Omega[\chi^{(1)},\chi^{(2)}]$ takes real values, is antisymmetric in $\chi^{(1)},\chi^{(2)}$ and is $\mathbb{R}$-bilinear (but not $\mathbb{C}$-bilinear).

\textit{Main properties of $\Omega$ (in the $q = 0$ subspace):}
Phonon deformations and type preserving ones are $\Omega$-orthogonal: To see this consider $\ket{\chi^{(1)}} = M \ket{\psi}$, with $M$ a constant $N \times N$ matrix and $\ket{\chi^{(2)}} = \ket{\psi'}$. Now
\begin{equation}
    \Omega \big[ M \ket{\psi},\lambda \ket{\psi'}\big] = i \int_{\rm{unit}\,\rm{cell}}\lambda^* \dfrac{\langle \psi'|(\mathbb{1}-P_{\psi})M| \psi \rangle}{\langle \psi | \psi \rangle} -
    \lambda \dfrac{\langle \psi| M^{\dagger}(\mathbb{1}-P_{\psi})| \psi' \rangle}{\langle \psi | \psi \rangle}
\end{equation}
where $\lambda \in \mathbb{C}$. Each of the two terms in the integrand is a derivative of a periodic function, so the integrand over a unit cell vanishes. 
For example, we have:
\begin{equation}
    \dfrac{\partial}{\partial z} \bigg( \dfrac{\langle \psi | M^{\dagger}|\psi \rangle}{\langle \psi | \psi \rangle}\bigg) =  \bigg( \dfrac{\langle \psi | M^{\dagger} (\mathbb{1} - P_{\psi})|\psi' \rangle}{\langle \psi | \psi \rangle}\bigg)
\end{equation}

The two phonon deformations are mutually conjugate: Consider $\ket{\chi^{(1)}} = w^{(1)} \ket{\psi'(z)}$ and $\ket{\chi^{(2)}} = w^{(2)} \ket{\psi'(z)}$ 
\begin{equation}
    \Omega \big(\chi^{(1)},\chi^{(2)}\big) = i \int_{\rm{unit}\,\rm{cell}} \dfrac{\langle \psi'|\mathbb{1}-P_{\psi}| \psi' \rangle}{\langle \psi | \psi \rangle} (w_1 w^*_2 - w_2 w^*_1)
\end{equation}
The integrand is equal to twice the topological charge density because $\ket{\psi}$ is holomorphic. This implies that
\begin{equation}
    \Omega \big(\chi^{(1)},\chi^{(2)}\big) = 4 \pi i N (w_1 w^*_2 - w_2 w^*_1)
    \label{topcomega}
\end{equation}
Let us now further investigate the symplectic form within this subspace of type-preserving deformations, i.e we take $\ket{\chi^{(1)}} = M_1 \ket{\psi}$ and $\ket{\chi^{(2)}} = M_2 \ket{\psi}$. We can then write 
\begin{equation}
    \Omega \big[ M_1 \ket{\psi},M_2 \ket{\psi}\big] = i \int_{\rm{unit}\,\rm{cell}} \dfrac{\langle \psi|M_2^{\dagger}(\mathbb{1}-P_{\psi})M_1| \psi \rangle}{\langle \psi | \psi \rangle} - \dfrac{\langle \psi| M_1^{\dagger}(\mathbb{1}-P_{\psi})M_2| \psi \rangle}{\langle \psi | \psi \rangle}
\end{equation}
Furthermore, any matrix $M$ can be written as a sum of hermitian and anti-hermitian matrices. If $M_1$ and $M_2$ are both hermitian/non-hermitian, then the corresponding deformations are $\Omega-$orthogonal. 

Matrices with antihermitian generators correspond to infinitesimal SU(N) transformations, they form a $N^2-1$ dimensional subspace over $\mathbb{R}$ playing the role of $q_j$ coordinates ($1 \leq j \leq N^2-1$). We also get a complementary $N^2-1$ dimensional subspace, composed of the traceless hermitian deformations, playing the role of $p_j$ (conjugate) coordinates. This yields the $N^2-1$ Jordan blocks because invariance of the total energy under $q-$translations imposes that the Taylor expansion around an optimal crystal contain only $p_ip_j$ terms.

For phonon deformations $\ket{\chi} = \lambda \ket{\psi'}$, we may write $\lambda = q_0 + ip_0$ with $q_0$ and $p_0$ real and eq.~(\ref{topcomega}) shows that (after a proper normalization) $q_0$ and $p_0$ can be taken as a pair of canonically conjugate coordinates. So phonon deformations (at $q = 0$) give rise to a zero frequency oscillator, instead of a Jordan block.

In the presence of anisotropies described by eq. (2) of the main text, the original
$SU(4)$ symmetry is broken into the much smaller two-dimensional $U(1)_P \times U(1)_S$ subgroup, turning 13 among the previous 15 Jordan blocks into finite
frequency oscillators. Moving away from the $q=0$ sector and considering small
but finite $q$ values also transforms each remaining Jordan block and the phonon
mode into an oscillator mode, whose frequency depends smoothly on $q$.

\end{document}